\documentclass[twocolumn,trackchanges]{aastex631}
\usepackage{amsmath,booktabs,amssymb}
\usepackage{multirow}
\usepackage[version=4]{mhchem}

\begin{document}

\title{Discovery of Interstellar 2-Cyanoindene (2-\ce{C9H7CN}) in GOTHAM Observations of TMC-1} 

\author{Madelyn L. Sita}
\altaffiliation{These authors contributed equally to this work.}
\affiliation{Department of Chemistry, University of Virginia, Charlottesville, VA 22904, USA}
\author{P. Bryan Changala}
\altaffiliation{These authors contributed equally to this work.}
\affiliation{Center for Astrophysics \textbar{} Harvard \& Smithsonian, Cambridge, MA 02138, USA}
\author{Ci Xue}
\affiliation{Department of Chemistry, Massachusetts Institute of Technology, Cambridge, MA 02139, USA}
\author{Andrew M. Burkhardt}
\affiliation{Department of Physics, Wellesley College, Wellesley, MA 02481, USA}
\author{Christopher N. Shingledecker}
\affiliation{Department of Physics and Astronomy, Benedictine College, Atchison, KS 66002, USA}
\author{Kin Long Kelvin Lee}
\affiliation{Accelerated Computing Systems and Graphics Group, Intel Corporation, 2111 NE 25th Ave, Hillsboro, OR 97124, USA}
\affiliation{Department of Chemistry, Massachusetts Institute of Technology, Cambridge, MA 02139, USA}
\affiliation{Center for Astrophysics \textbar{} Harvard \& Smithsonian, Cambridge, MA 02138, USA}
\author{Ryan A. Loomis}
\affiliation{National Radio Astronomy Observatory, Charlottesville, VA 22903, USA}
\author{Emmanuel Momjian}
\affiliation{National Radio Astronomy Observatory, Socorro, NM 87801, USA}
\author{Mark A. Siebert}
\affiliation{Department of Astronomy, University of Virginia, Charlottesville, VA 22904, USA}
\author{{Divita Gupta}}
\affiliation{{I. Physikalisches Institut, Universit\"at zu K\"oln, Z\"ulpicher Str. 77, 50937 K\"oln, Germany}}
\author{Eric Herbst}
\affiliation{Department of Chemistry, University of Virginia, Charlottesville, VA 22904, USA}
\affiliation{Department of Astronomy, University of Virginia, Charlottesville, VA 22904, USA}
\author{Anthony J. Remijan}
\affiliation{National Radio Astronomy Observatory, Charlottesville, VA 22903, USA}
\author{Michael C. McCarthy}
\affiliation{Center for Astrophysics \textbar{} Harvard \& Smithsonian, Cambridge, MA 02138, USA}
\author{Ilsa R. Cooke}
\affiliation{Department of Chemistry, University of British Columbia, 2036 Main Mall, Vancouver BC V6T 1Z1, Canada}
\author[0000-0003-1254-4817]{Brett A. McGuire}
\affiliation{Department of Chemistry, Massachusetts Institute of Technology, Cambridge, MA 02139, USA}
\affiliation{National Radio Astronomy Observatory, Charlottesville, VA 22903, USA}
\affiliation{Center for Astrophysics \textbar{} Harvard \& Smithsonian, Cambridge, MA 02138, USA}

\correspondingauthor{Brett A. McGuire}
\email{brettmc@mit.edu}

\begin{abstract}

We present laboratory rotational spectroscopy of five isomers of cyanoindene (2-, 4-, 5-, 6-, and 7-cyanoindene) using a cavity Fourier-transform microwave spectrometer operating between 6--40\,GHz.  Based on these measurements, we report the detection of 2-cyanoindene (1H-indene-2-carbonitrile; 2-\ce{C9H7CN}) in GOTHAM line survey observations of the dark molecular cloud TMC-1 using the Green Bank Telescope at centimeter wavelengths.  Using a combination of Markov Chain Monte Carlo (MCMC), spectral stacking, and matched filtering techniques, we find evidence for the presence of this molecule at the 6.3$\sigma$ level.  This provides the first direct observation of the ratio of a cyano-substituted polycyclic aromatic hydrocarbon (PAH) to its pure hydrocarbon counterpart, in this case indene, in the same source.  We discuss the possible formation chemistry of this species, including why we have only detected one of the isomers in TMC-1.  We then examine the overall hydrocarbon:CN-substituted ratio across this and other simpler species, as well as {compare} to those ratios predicted by astrochemical models.  We conclude that while astrochemical models are not yet sufficiently accurate to reproduce absolute abundances of these species, they do a good job at predicting the ratios of hydrocarbon:CN-substituted species, further solidifying -CN tagged species as excellent proxies for their fully-symmetric counterparts.

\end{abstract}

\keywords{}

\section{Introduction}

Polycyclic aromatic hydrocarbons (PAHs) may account for up to a quarter of the total carbon mass of the interstellar medium (ISM)~\citep{Tielens:2008:289}. Despite their apparent ubiquity, the formation and evolution of PAHs remain poorly {constrained, highlighting the need for sensitive,} targeted observations of individual PAH species in a variety of astronomical environments. Although such specificity is readily achieved by radio frequency measurements of pure rotational spectra, PAHs are notoriously difficult, if not impossible, to detect by this approach because they are often highly symmetric and therefore possess no permanent dipole moment, or are at best weakly polar when not symmetric. 

However, replacing even a single hydrogen on a ``pure" (i.e. containing only carbon and hydrogen) PAH with a polar functional group, such as the nitrile (or cyano) unit, $-$C$\equiv$N, yields a spectroscopically bright surrogate. In fact, the first detections of specific interstellar PAHs were nitrile derivatives of naphthalene, 1- and 2-cyanonaphthalene (\ce{C10H7CN}; \citealt{McGuire:2021:1265}).  These discoveries were made in the dark, cold Taurus Molecular Cloud 1 (TMC-1), the same source where the simplest aromatic nitrile benzonitrile (\ce{C6H5CN}) was first detected \citep{McGuire:2018:202}.  {Taken together, and combined with complementary results from the QUIJOTE project observing TMC-1 as well (see, e.g., \citealt{Cernicharo:2021:L15}), we are seeing an entirely new dimension \citep{McCarthy:2021:3231}} to the already rich inventory of complex organic  molecules in this source~\citep{Gratier:2016:25}. Laboratory studies indicate that nitrile functionalization of benzene occurs facilely via a direct reaction of CN radicals with the aromatic double bonds, suggesting that  CN-tagging has the potential to be a broadly applicable scheme/strategy for quantifying otherwise radio-dark PAHs~\citep{Lee:2019:2946}. Because benzene and naphthalene lack a permanent dipole moment, however,  it is not possible constrain the PAH-nitrile chemistry occurring in astronomical sources.

\begin{figure*}
\centering
\includegraphics[width=0.8\textwidth]{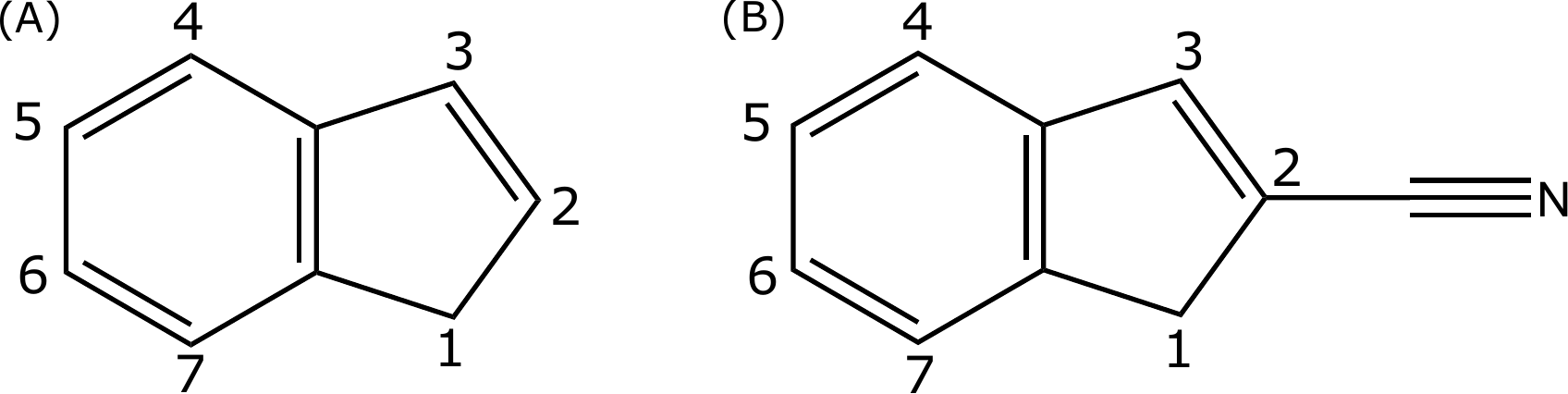}
\caption{\label{fig:indene} The chemical structure of indene (A) and 2-cyanoindene (B; 2-\ce{C9H7CN}). Sites 1 -- 7 label the nitrile substitution positions of the seven possible isomers of cyanoindene.} 
\end{figure*}

The recent discovery of indene (\ce{C9H8}, Fig.~\ref{fig:indene}) in TMC-1, the first interstellar detection of a pure PAH, now affords one an opportunity to directly link pure PAHs and their functionalized nitrile counterparts~{\citep{Burkhardt:2021:L18,Cernicharo:2021:L15}}, provided the latter can also be detected in the same source. To this end, this paper reports new laboratory measurements of the radio frequency transitions of several cyanoindene (\ce{C9H7CN}) isomers in the 6--40~GHz spectral band. These hyperfine-resolved measurements provide the high-precision spectroscopic parameters necessary for astronomical searches of cyanoindene in cold molecular clouds. We then use these new measurements in conjunction with the latest observational data from the GOTHAM (Green Bank Telescope Observations of TMC-1: Hunting for Aromatic Molecules) survey to search for the cyanoindenes toward TMC-1. We detect the 2-cyanoindene isomer (1H-indene-2-carbonitrile; 2-\ce{C9H7CN}) and quantify upper limits on the column densities of four other cyanoindene isomers. We discuss these results in the context of broader astrochemical modeling of TMC-1, focusing on the ability of cyano-substituted molecules to serve as proxies of their pure hydrocarbon counterparts.

%

\section{Laboratory Measurements}

The cyanoindenes were prepared in the laboratory using similar methods to our previous experiments with benzonitrile~\citep{Lee:2019:2946} and cyanocyclopentadiene~\citep{McCarthy:2021:176,Lee:2021:L2}. 
A 1-mL sample of indene, absorbed onto a cotton swab, was placed in-line behind a pulsed solenoid valve, backed with a dilute (0.1\%) mixture of acetonitrile, CH$_3$CN, in neon at a pressure of 2.5~kTorr. 
The mixture was supersonically expanded along the axis of a cavity Fourier transform microwave (FTMW) spectrometer~\citep{Grabow:2005:093106} in 400 $\mu$s gas pulses at a rate of 6~Hz. 
During each gas pulse, a 1.2 kV discharge was struck between two copper electrodes placed immediately after the valve aperture producing reactive CN radicals, which further react with indene to produce cyanoindene isomers.
The molecules are efficiently cooled in the supersonic expansion to an internal rotational temperature of ca. 2~K.

Using a combination of our previous indene measurements~\citep{Burkhardt:2021:L18} with theoretical structural calculations of the CN-derivatives, we predicted the rotational transition frequencies of all seven possible cyanoindene isomers over the 6--40~GHz range of the cavity-FTMW spectrometer.  Owing to their large dipole moments (ca. 5~D), transitions from several isomers were readily observed after brief searches.  The rotational constants and dipole moments of each of the isomers obtained at the $\omega$B97X-D/6-31+G(d) level of theory and basis set using the Gaussian 16 suite of programs \citep{Gaussian16} are given in Table~\ref{tab:calculations}, along with relative energies obtained with the G3//B3LYP method.  Ultimately, we detected five isomers with comparable abundances (2-, 4-, 5-, 6-, and 7-cyanoindene; see Fig.~\ref{fig:indene} for the labeling convention), with the number corresponding to the site of the CN group. The method and basis are in line with calculations performed in our earlier GOTHAM papers, with well-characterized uncertainty in the rotational constants and dipole moments \citep{Lee:2020:898}. The raw line list for each is available in the supplementary material.

The spectroscopic constants of the observed isomers were determined by least-squares fitting to an A-reduced ($I^r$) effective Hamiltonian including quartic centrifugal distortion and $^{14}$N-quadrupole hyperfine parameters. These constants are summarized in Table~\ref{tab:spec}. Generally, the fit quality is quite good, with residuals comparable to the 2~kHz measurement uncertainty. Some isolated centrifugal distortion constants (e.g. $\Delta_{JK}$, $\Delta_K$, and $\delta_K$ of isomers 5 and 6) are poorly determined, but were included in the fit to treat {all isomers} uniformly. They have little effect on the dominant $a$-type transitions of these isomers over this frequency range. The similarity of the rotational and hyperfine constants for the isomer pairs 4/7 and 5/6 demonstrates the near-structural equivalence of their respective substitution sites.

\begin{table*}[]
    \centering
    \caption{Calculated equilibrium spectroscopic constants (MHz), dipole moment (Debye), and relative energy (kJ/mol) for each possible cyanoindene isomer. Dipole moments have a nominal uncertainty of $\pm0.5$ Debye \citep{Lee:2020:898}.}    
    \begin{tabular}{c r r r r r r r}
    \toprule
    Parameter   &   1           &   2           &   3           &   4           &   5           &   6           &   7           \\
    \midrule
    $A$         &   2126.490335 &   3770.591920 &   2148.858711 &   1732.450016 &   3264.72881  &   3252.621963 &   1723.113358 \\
    $B$         &   971.888905  &   676.199976  &   953.096847  &   1145.202556 &   727.193886  &   728.881556  &   1155.797670 \\
    $C$         &   710.942917  &   575.424855  &   662.960551  &   692.411160  &   596.922047  &   597.652000  &   694.743761  \\
    &\\
    $\mu_a$     &   3.02        &   4.75        &   3.70        &   4.99        &   5.76        &   5.62        &   4.09        \\
    $\mu_b$     &   2.21        &   0.29        &   3.19        &   2.08        &   1.31        &   0.43        &   1.61        \\
    $\mu_c$     &   1.87        &   --          &   --          &   --          &   --          &   --          &   --          \\
    &\\
    $E$           &   26.88       &   0.00        &   4.02        &   3.40        &   5.98        &   4.58        &   3.43        \\
    \bottomrule
    \end{tabular}
    \label{tab:calculations}
\end{table*}

\begin{table*}[t]
\caption{\label{tab:spec} The fitted spectroscopic constants of the observed cyanoindene isomers. The microwave transition frequencies were fitted to an A-reduced (I$^r$) effective Hamiltonian including
quartic centrifugal distortion terms and $^{14}$N nuclear quadrupole hyperfine structure. All dimensionful parameters are reported in MHz.  }
\begin{tabular}{ccccccc}
Parameter & 2 & 4 & 5 & 6 & 7 \\ 
\hline 

$A$ & 3754.427(22) & 1728.20692(42) & 3254.849(10) & 3241.0526(89) & 1718.76438(50)\\
$B$ & 675.99176(12) & 1145.625094(76) & 726.94511(16) & 728.82948(16) & 1156.25104(11)\\
$C$ & 574.98249(12) & 691.881285(51) & 596.49419(13) & 597.29716(12) & 694.190200(64)\\
\\
$\Delta_J \times 10^6$ & $5.84(22)$ & $67.65(33)$ & $8.31(33)$ & $8.52(34)$ & $71.29(86)$\\
$\Delta_{JK} \times 10^6$ & $110(12)$ & $-202.9(32)$ & $4.4(69)$ & $7.9(70)$ & $-213.3(58)$\\
$\Delta_K \times 10^3$ & $-18.6(73)$ & $0.315(14)$ & $0.14(313)$ & $-2.0(26)$ & $0.312(24)$\\
$\delta_J \times 10^6$ & $1.42(21)$ & $30.01(17)$ & $1.94(21)$ & $2.09(13)$ & $31.71(47)$\\
$\delta_K \times 10^3$ & $-0.105(58)$ &$0.0420(27)$ & $0.042(50)$ & $0.005(50)$ & $0.0403(41)$\\
\\
$\chi_{aa}$ & $-4.223(35)$ & $-2.0873(56)$ & $-3.884(29)$ & $-3.853(33)$ & $-2.0864(56)$\\
$\chi_{bb}$ & $2.254(29)$ & $0.1927(50)$ & $1.941(18)$ & $1.971(19)$ & $0.1425(55)$\\ 
\\
$N_\text{lines}$$^a$ & 111 & 220 & 125 & 140 & 149\\
$\sigma_\text{norm}$$^b$ & $0.75$ & $0.74$ & $0.98$ & $0.73$ & $0.96$ \\
$(J,K_a)_\text{max}$$^c$ & $(18, 2)$ & $(12,6)$ & $(16,3)$ & $(18,3)$ & $(10,5)$ \\
\hline 
\multicolumn{6}{l}{$^a$ The number of hyperfine-resolved transitions included in the fit.}\\
\multicolumn{6}{l}{$^b$ The root-mean-square value of the fit residuals normalized by the 2~kHz measurement uncertainty.}\\
\multicolumn{6}{l}{$^c$ The maximum value of the $J$ and $K_a$ rotational quantum numbers of the states included in the fit.}\\
\multicolumn{6}{l}{{\emph{Note --} Uncertainties are provided in parentheses in units of the last significant digit.}}\\
\end{tabular}
\end{table*}

\section{Observations}

The fourth data reduction of the GOTHAM collaboration survey, hereafter referred to as DR4, includes new observations taken between June 2020 and May 2022 on the Robert C. Byrd 100 m Green Bank Telescope (GBT) in Green Bank, West Virginia. These new observations include the remainder of project code AGBT19B-047 not included in previous releases as well as AGBT20A-516, AGBT21A-414, and AGBT21B-210 which contributed an additional 953 hours to those presented in DR3 \citep{Barnum:2022:2716}. All previous inclusions of 19B-047 and 18B-007 were reduced again to maintain a consistent method across the new data set, while the archived data from project codes AGBT18A-333, AGBT17A-164 and AGBT17A-43 were folded in for the final spectra. DR4 extends the frequency coverage of the survey to 7.906-36.411 GHz (24.9 GHz total observed bandwidth) with a few gaps, {at a frequency resolution of 1.4\,kHz corresponding to $\sim$0.01-0.05\,km\,s$^{-1}$ across the full bandwidth}, and improves the sensitivity (typically $\sim$2--5\,mK RMS depending on ultimate integration time at any given frequency) and radio frequency interference (RFI) removal in areas already covered by previous reductions. This is, in part, due to a more consistent and precise method of data reduction involving a careful investigation at each observing session, scan by scan, to identify and remove RFI artifacts.  {The GBT beam varies from $\sim$22--95$^{\prime\prime}$ over the covered frequency range.}

The target, as in all GOTHAM observations, was the cyanopolyyne peak of TMC-1 at (J2000) $\alpha$~=~04$^h$41$^m$42.50$^s$ $\delta$~=~+25$^{\circ}$41$^{\prime}$26.8$^{\prime\prime}$. All observations were conducted using position-switching mode (ON-OFF) in which the off position was 1\textdegree\ off target and was confirmed to be clear of emission. Every $\sim$2 hours, pointing and focus observations were taken using the calibrators J0530+1331 and J0359+5057, which have also been used for flux density scale calibration. Typical pointing solutions converged to $\leq$5$^{\prime\prime}$.

{The GBT receivers are primarily calibrated by means of an internal noise diode, which we assume gives an absolute flux density calibration uncertainty of, at best, $\sim$30\%.  The noise diode in the X-band receiver was calibrated as recently as 2018 and referenced to Karl G. Jansky Very Large Array (VLA) flux density measurements (see \url{http://www.gb.nrao.edu/GBTCAL/}), and is therefore assumed to be better than 30\%.  We have taken several steps to improve both the absolute flux calibration of the remaining measurements, and to ensure relative agreement between the two (and with X-band).}  Initial flux density scale calibration was performed on data taken through 2020 by comparing to VLA flux density measurements obtained in 2019 and described in detail in \citet{McGuire:2020:L10}.  In an effort to monitor these sources for changes in flux density, and to cover additional frequency ranges not yet observed in the GOTHAM project as of the original calibration in 2020, VLA measurements were taken in September 2021 for both J0359+5057 and J0530+1331.  Flux densities were obtained between 14--16\,GHz (corresponding to new GOTHAM K$_u$-band measurements), 18.1--20.8\,GHz (corresponding to new GOTHAM K-band measurements), and 34.1--36.0,\,GHz (corresponding to new GOTHAM Ka-band measurements).  Based on a comparison to observations of these sources with the GBT, we found agreement at the 5--10\% level averaged across all observing sessions, indicating no major adjustments were needed.  

As noted in \citet{McGuire:2020:L10}, however, given short-term variability of the sources, we still estimate the uncertainty in our flux density scale calibration is at best $\sim$20\%. {We would note that each of the bands is self-consistently calibrated with an internal noise diode at the GBT.  Thus, within a band, we believe that whatever the actual error in the flux density scale is (which we believe to be $\lesssim$20\% as discussed above), it is likely a uniform error across the entire band.  Between bands, however, we suspect that the errors are non-uniform. These band-to-band variations will be more rigorously addressed with the final data reduction of the completed survey.}

\section{Observational Analysis}

In order to derive physical parameters (column density [$N_T$], excitation temperature [$T_{ex}$], linewidth [$\Delta V$], and source size [$^{\prime\prime}$]) for the target molecules in our observations, we used the same Markov-Chain Monte Carlo (MCMC) model employed in prior GOTHAM analyses (see, e.g., \citealt{Siebert:2022:21,Lee:2021:L11}). {The technique is described in detail in \citet{Loomis:2021:188}, and an exhaustive analysis of potential sources of errors, uncertainties, and spurious signals is described in the Supplementary Information for \citep{McGuire:2021:1265}}.  In short, the MCMC model calculates probability distributions and co-variances for these parameters which are used to describe the emission of molecules observed in our data.  The resulting corner plots for our target species are shown in the Appendix (Figs.~\ref{fig:indene_corner} \& \ref{fig:indene_cn_corner}). We adopt the 50$^{th}$ percentile value of the posterior probability distributions as the representative value of each parameter for the molecule, and use the 16$^{th}$ and 84$^{th}$ percentile values for the uncertainties.  For probabilities that show a Gaussian distribution, these correspond to the 1$\sigma$ uncertainty level. Many of our resulting probability distributions are indeed either Gaussian or nearly Gaussian, and thus these values are usually quite representative of the 1$\sigma$ uncertainties.  One of the advantages of the MCMC technique over a traditional least-squares fit approach, however, is that far more of parameter space is explored.  Correspondingly, a much larger exploration of the uncertainty space is performed as well, including highlighting parameters that may be highly co-variant with one another.  This manifests as non-separable distributions in the corner plots.

To explore this parameter space with our MCMC approach, a model of the molecular emission is generated for each set of parameters using the \texttt{molsim} software package \citep{molsim} and following the conventions of \citet{Turner:1991:617} for a single excitation temperature and accounting for the effect of optical depth.  Prior observations both from GOTHAM \citep{Xue:2020:L9} and others \citep{Dobashi:2018:82,Dobashi:2019:88} have found that most emission seen at cm-wavelengths in TMC-1 can be separated into contributions from four distinct velocity components within the larger structure, at approximately 5.4, 5.6, 5.8, and 6.0\,km\,s$^{-1}$ \citep{Loomis:2021:188}.  In some cases, especially for less-abundant species where there is not a clear detection in one of the velocity components, we find that a three-component model has performed better \citep{McGuire:2020:L10}.

We began first by re-visiting the original detection of \ce{C9H8} from \citet{Burkhardt:2021:L18} which used the DR2 data from GOTHAM.  Using the DR4 data, we find a three-component model performs best with the more sensitive DR4 data, with strong detections in the 5.6, 5.8, and 6.0\,km\,s$^{-1}$ components.  These v$_{lsr}$ values, along with the value for $T_{ex}$, were used as priors for the subsequent analysis of the \ce{C9H7CN} isomers, following the same procedure.  

To determine the statistical evidence that our model of the emission of these molecules is consistent with the data, we followed the procedures described in detail in \citet{Loomis:2021:188} and performed a spectral stack and matched filtering analysis.  Briefly, a weighted average of the observational spectra in velocity space and centered on each spectral line of a target molecule was performed.  The weights were determined by the relative intensity of the expected emission (based on the MCMC-derived parameters) and the local RMS noise of the observations. Considering the weak expected intensities for both \ce{C9H8} and the \ce{C9H7CN} isomers, any observational windows containing emission at $>$5$\sigma$ were ignored in this analysis.  {Given the low line-density of the spectra, the number of spectral regions flagged this way is small; e.g., for indene, 1\% of the data needed to be rejected in this way, while for 2-cyanoindene, only 0.6\% of the data were rejected.}  Simulated spectra of the molecular emission using the same MCMC-derived parameters were then also generated and stacked using identical weights.  This simulation was then used as a matched filter which is passed through the observational signal.  The resulting impulse response function represents the statistical evidence that our model of the emission from the molecule -- and thus our derived parameters for the molecule -- is consistent with the observations.  In addition to the details of the methodology provided in \citet{Loomis:2021:188}, the appendices of \citet{McGuire:2021:1265} include an extensive analysis of the robustness of the methodology, including the improbability of spurious signals and the minimal impact of red-noise on the procedure.

\section{Results}

The resulting parameters from the MCMC inference to \ce{C9H8} emission in the DR4 observational data are shown in Table~\ref{tab:indene_vals}; the corner plot is shown in Fig.~\ref{fig:indene_corner}.   The new column density derived from the DR4 data, $9.04^{+0.96}_{-0.96}\times 10^{12}$ cm$^{-2}$, falls firmly within the uncertainty range of our previous value, $9.60^{+4.33}_{-1.57}\times 10^{12}$ cm$^{-2}$ \citep{Burkhardt:2021:L18}, but itself has a narrower range of uncertainties reflecting the increased quality of the DR4 data.  We note that the value derived for the excitation temperature, $T_{ex} = 5.64$\,K, is lower than that found previously in our DR2 data of $T_{ex} = 8.55$\,K.  We and others (see, e.g., \citealt{Cernicharo:2021:L9}) continue to find excitation {temperatures} within TMC-1 fall between $\sim$5--9\,K. In the GOTHAM data, the values derived are extremely sensitive to the number of higher-energy transitions observable at higher frequencies, the quality of the data at those frequencies, and the relative flux calibration throughout the survey.  This is further complicated by the large uncertainties in the source sizes, which differentially affect beam dilution corrections at the low ($\sim$10\,GHz) and high ($\sim$30\,GHz) ends of our data, where the telescope beam sizes vary significantly. Combined with the difficulties in constraining the flux uncertainty discussed above, we would caution that the quoted statistical uncertainties on derived values of $T_{ex}$ in GOTHAM data are likely underestimated. The ultimate effect of the small {change} in $T_{ex}$ on the other derived parameters, particularly $N_T$, however, is negligible.

\begin{figure*}[bt]
    \centering
    \includegraphics[width=0.49\textwidth]{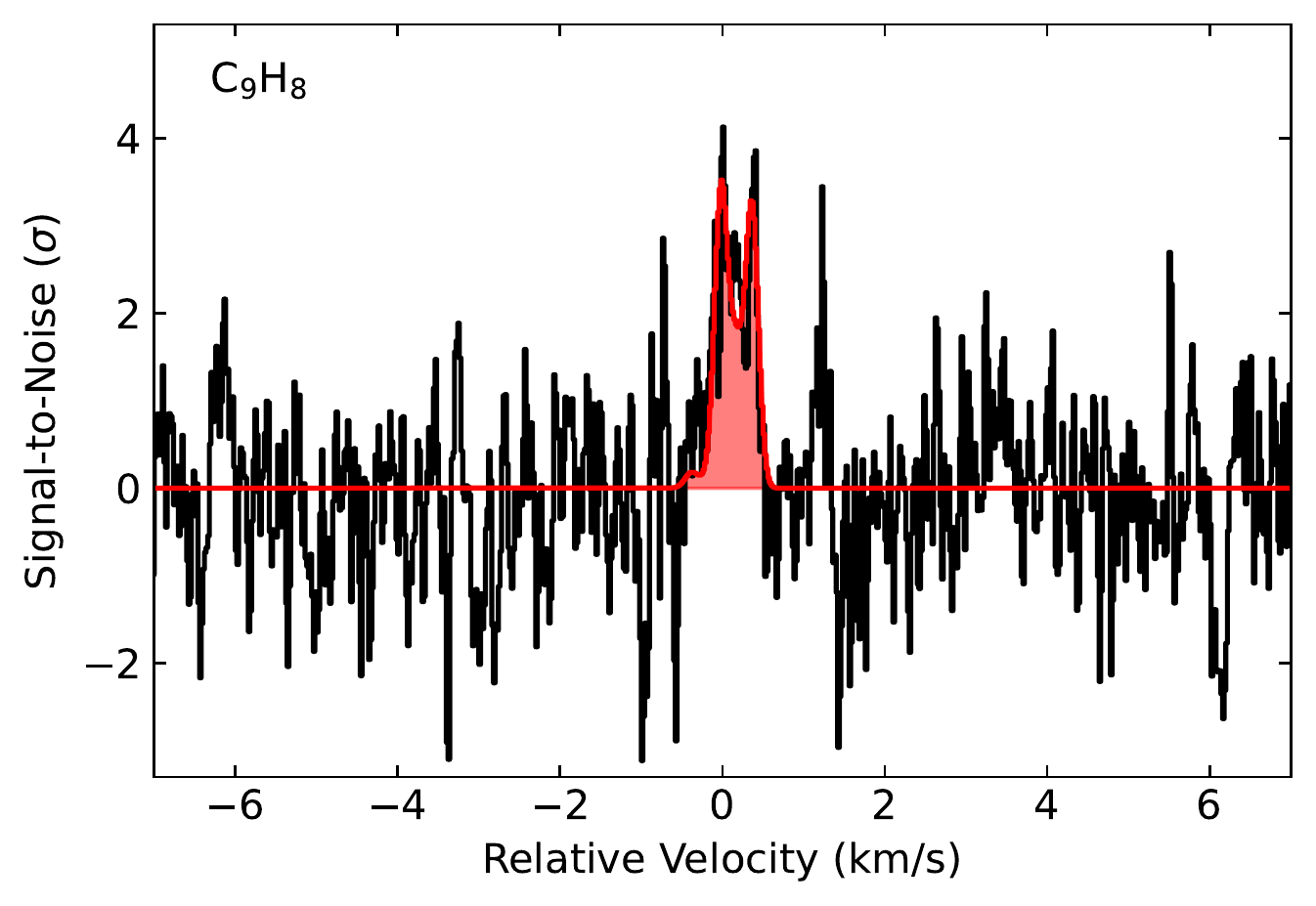}
    \includegraphics[width=0.49\textwidth]{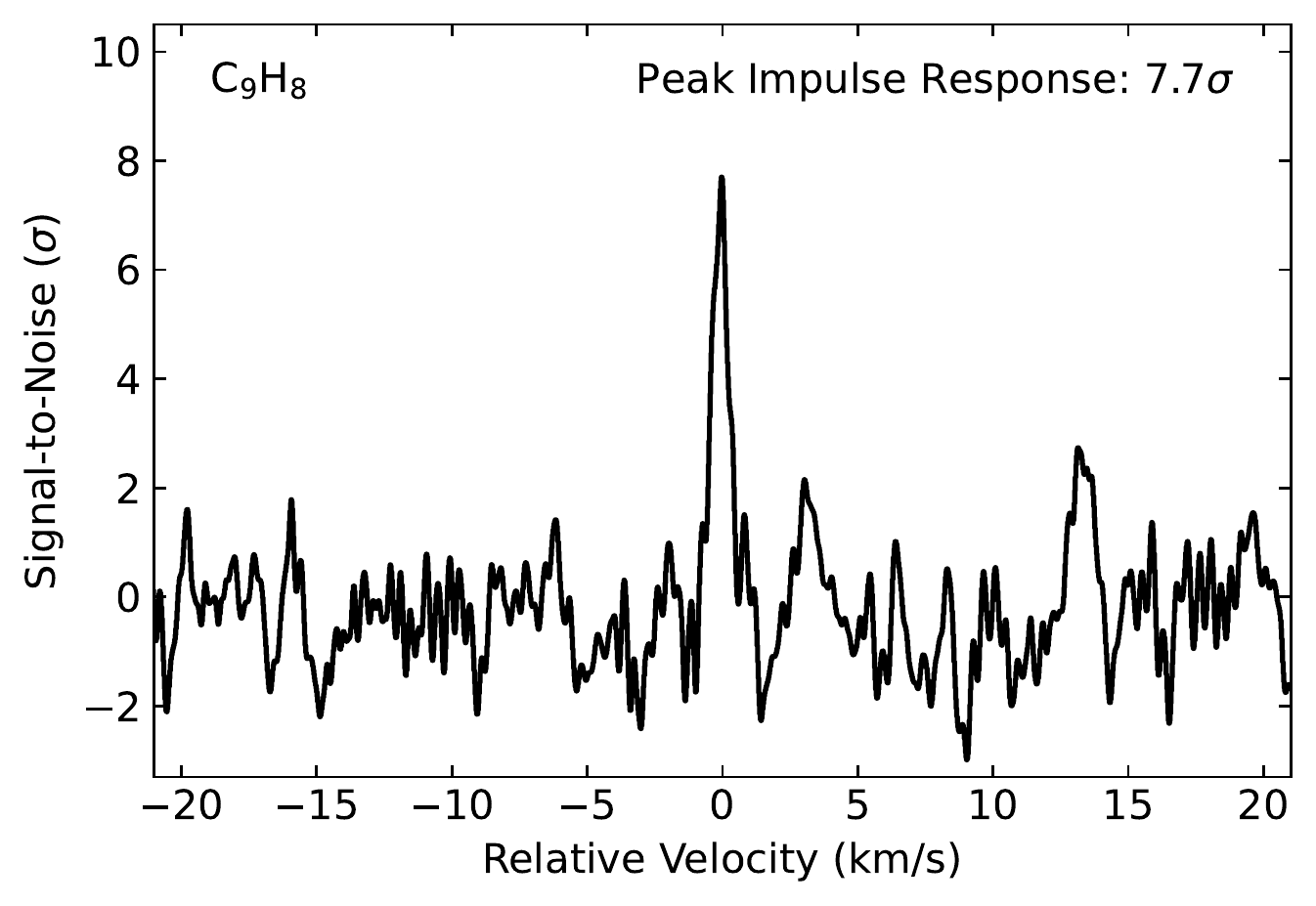}
    \caption{Velocity stacked and matched filter spectra of \ce{C9H8}. The intensity scales are the signal-to-noise ratios (SNR) of the response functions when centered at a given velocity. The ``zero'' velocity corresponds to the channel with the highest intensity to account for blended spectroscopic transitions and variations in velocity component source sizes. (\emph{Left}) The stacked spectra from the GOTHAM DR4 data are displayed in black, overlaid with the expected line profile in red from our MCMC fit to the data.  The signal-to-noise ratio is on a per-channel basis. (\emph{Right}) Matched filter response obtained from cross-correlating the simulated and observed velocity stacks in the left panel; value annotated corresponds to the peak impulse response of the matched filter.}
    \label{fig:indene_stack}
\end{figure*}

Using these updated parameters for our 3-component model of the emission of \ce{C9H8}, we generated a spectral stack and performed a matched filtering analysis as describe above.  The results are shown in Fig.~\ref{fig:indene_stack}.  The evidence for a detection of indene in our data has increased from the 5.7$\sigma$ as reported in \citet{Burkhardt:2021:L18} to 7.7$\sigma$, reflecting the increased spectral coverage and higher signal-to-noise of the DR4 dataset. 

\begin{table}[h!]
    \centering
        \caption{Summary Statistics of the Marginalized \ce{C9H8} Posterior}
    \begin{tabular}{ c c c c c}
    \toprule
		$v_{lsr}$	&	Size	&	$N_T$	&	$T_{ex}$	&	$\Delta V$	\\
		(km\,s$^{-1}$)	&	($^{\prime\prime}$)	&	(10$^{12}$cm$^{-2}$)	&	(K)	&	(km\,s$^{-1}$)\\
	\midrule
		$5.638^{+0.013}_{-0.011}$	 & 	$423^{+190}_{-204}$	 & 	$3.69^{+0.53}_{-0.45}$	 & 	 \multirow{3}{*}{$5.64^{+0.78}_{-0.65}$}	 & 	 \multirow{3}{*}{$0.187^{+0.028}_{-0.022}$}\\
		$5.835^{+0.034}_{-0.028}$	 & 	$387^{+210}_{-228}$	 & 	$1.67^{+0.53}_{-0.71}$	 & 	 & 	 \\
		$6.019^{+0.015}_{-0.012}$	 & 	$430^{+184}_{-204}$	 & 	$3.67^{+0.60}_{-0.47}$	 & 	 & 	 \\
	\midrule
		\multicolumn{5}{c}{$N_T$(Total): $9.04^{+0.96}_{-0.96}\times 10^{12}$ cm$^{-2}$}\\
    \bottomrule 
    \end{tabular}
        \begin{minipage}{\columnwidth}
        	\footnotesize
        	{Note} -- The quoted uncertainties represent the 16$^{th}$ and 84$^{th}$ percentile ($1\sigma$ for a Gaussian distribution) uncertainties.\\
        	$^{\dagger}$Uncertainties derived by adding the uncertainties of the individual components in quadrature.
        \end{minipage}     
    \label{tab:indene_vals}
\end{table}

\begin{figure*}[bt]
    \centering
    \includegraphics[width=0.49\textwidth]{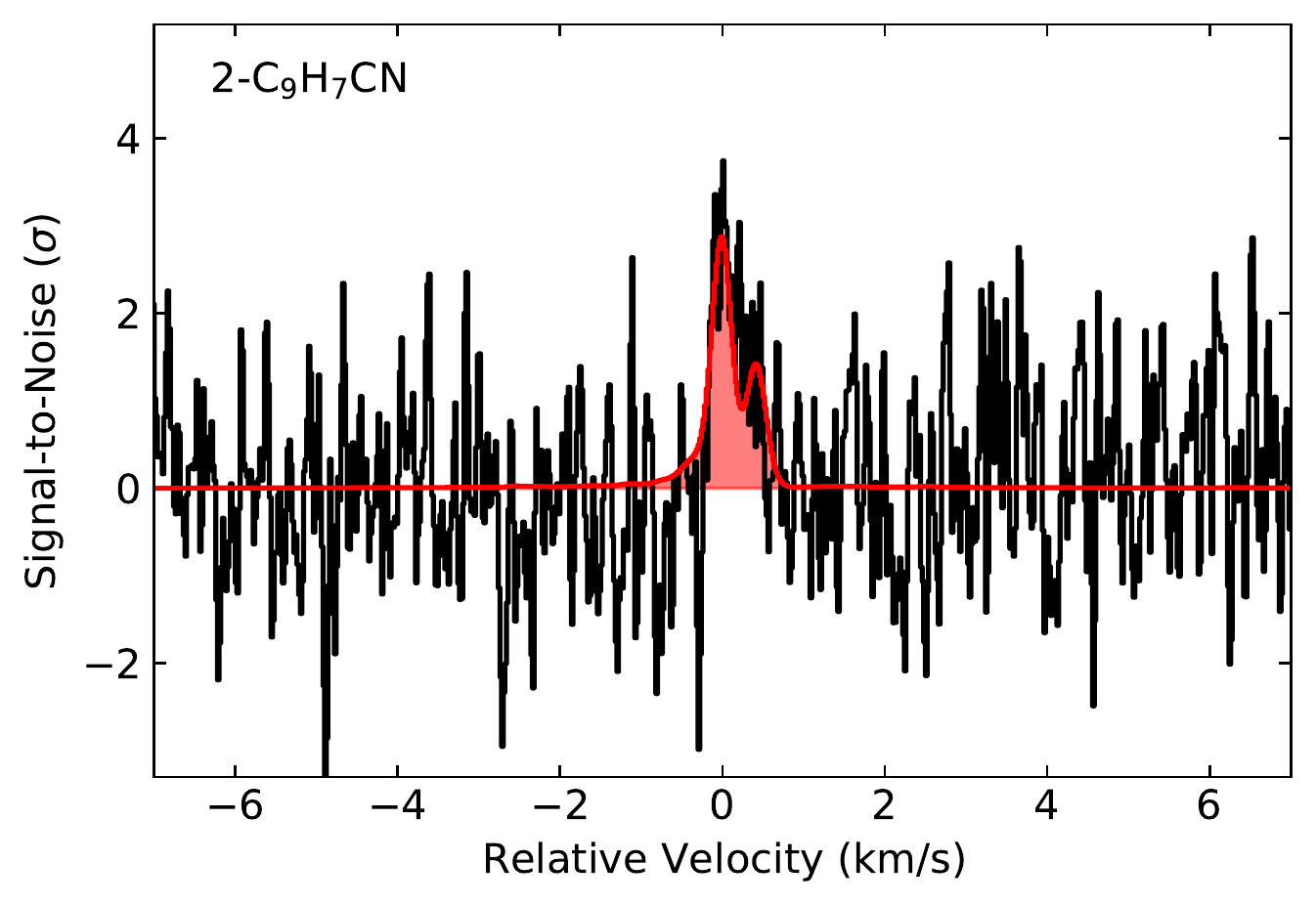}
    \includegraphics[width=0.49\textwidth]{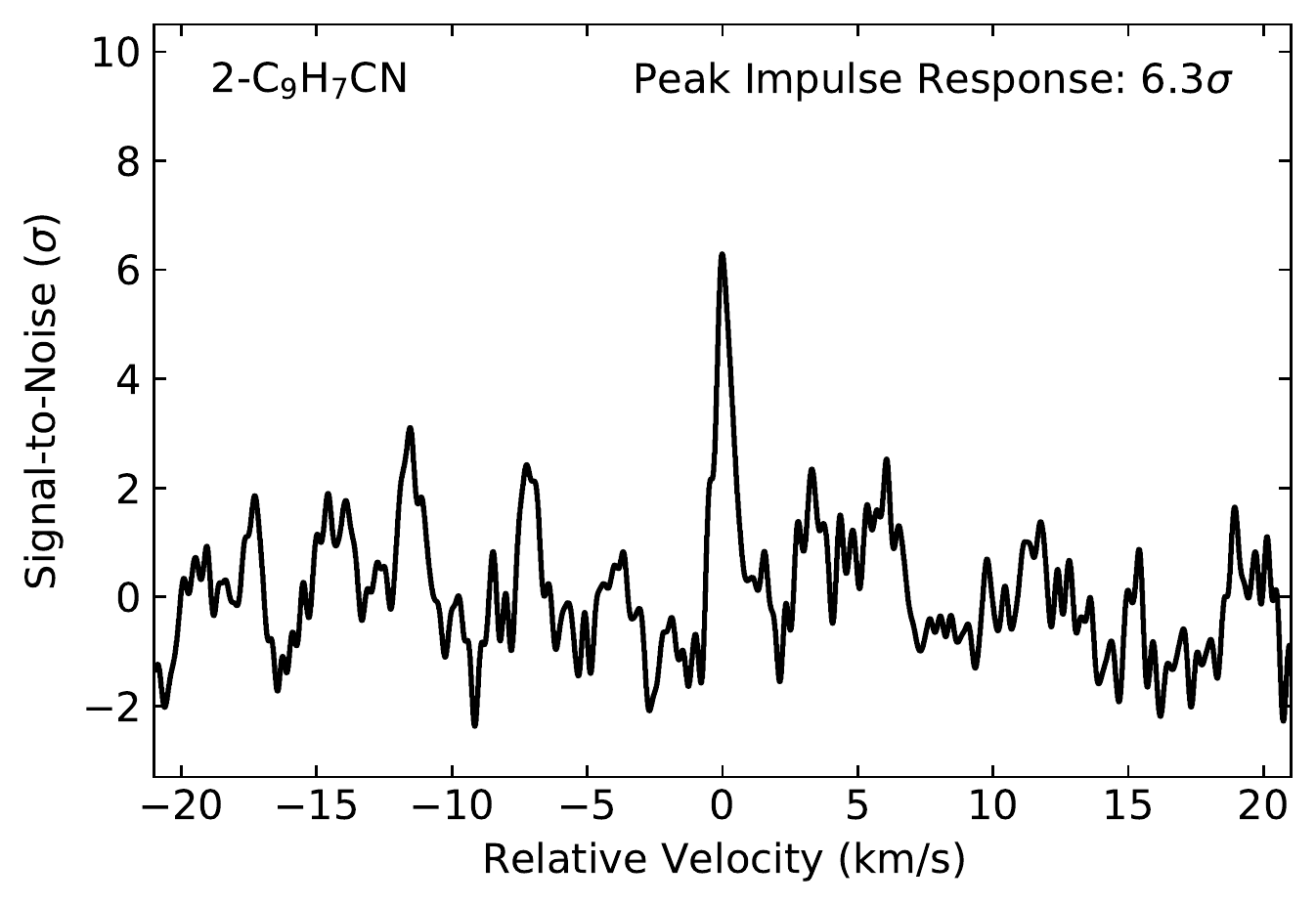}
    \caption{{Same as for Fig.~\ref{fig:indene_stack}} but for 2-\ce{C9H7CN} instead.}
    \label{fig:indene_cn_stack}
\end{figure*}

We see statistically significant evidence ($>$5$\sigma$) for the presence of only a single isomer of \ce{C9H7CN}: Isomer 2 (2-\ce{C9H7CN}), which exhibits a peak impulse response of 6.3$\sigma$.  The best-fit parameters from the MCMC fit for this isomer are shown in Table~\ref{tab:isomerB_vals}, and the spectral stack and matched filter are shown in Fig.~\ref{fig:indene_cn_stack}; the corner plot is shown in Fig.~\ref{fig:indene_cn_corner}.  MCMC analyses for the other isomers resulted in no significant signal being detected.  Upper limits to the column densities for the other species are provided in Table~\ref{tab:ncols}. 

\begin{table}[h!]
    \centering
        \caption{Summary Statistics of the Marginalized 2-\ce{C9H7CN} Posterior}
    \begin{tabular}{ c c c c c}
    \toprule
		$v_{lsr}$	&	Size	&	$N_T$	&	$T_{ex}$	&	$\Delta V$	\\
		(km\,s$^{-1}$)	&	($^{\prime\prime}$)	&	(10$^{10}$cm$^{-2}$)	&	(K)	&	(km\,s$^{-1}$)\\
	\midrule
		$5.492^{+0.047}_{-0.054}$	 & 	$435^{+179}_{-204}$	 & 	$6.85^{+2.08}_{-1.80}$	 & 	 \multirow{3}{*}{$5.09^{+0.52}_{-0.48}$}	 & 	 \multirow{3}{*}{$0.247^{+0.048}_{-0.047}$}\\
		$5.899^{+0.040}_{-0.168}$	 & 	$394^{+206}_{-227}$	 & 	$6.04^{+3.99}_{-2.18}$	 & 	 & 	 \\
		$5.946^{+0.031}_{-0.027}$	 & 	$380^{+220}_{-230}$	 & 	$8.10^{+3.96}_{-3.61}$	 & 	 & 	 \\
	\midrule
		\multicolumn{5}{c}{$N_T$(Total): $2.10^{+0.60}_{-0.46}\times 10^{11}$ cm$^{-2}$}\\
    \bottomrule 
    \end{tabular}
        \begin{minipage}{\columnwidth}
        	\footnotesize
        	{Note} -- The quoted uncertainties represent the 16$^{th}$ and 84$^{th}$ percentile ($1\sigma$ for a Gaussian distribution) uncertainties.\\
        	$^{\dagger}$Uncertainties derived by adding the uncertainties of the individual components in quadrature.
        \end{minipage}     
    \label{tab:isomerB_vals}
\end{table}

\begin{table}[t!]
    \centering
    \caption{Column density and column density upper limit determinations for the seven isomers of cyanoindene.}    
    \begin{tabular*}{\columnwidth}{c @{\extracolsep{\fill}} c}
    \toprule
    Isomer  &   Total Column Density                    \\
            &   ($10^{11}$ cm$^{-2}$)                   \\
    \midrule
    1       &   --$^{\dagger}$                          \\
    2       &   $2.10^{+0.60}_{-0.46}$    \\
    3       &   --$^{\dagger}$                          \\
    4       &   $<1.7$                     \\
    5       &   $<5.1$                     \\
    6       &   $<3.3$                     \\
    7       &   $<4.7$                     \\
    \bottomrule
    \end{tabular*}
        \begin{minipage}{\columnwidth}
        	\footnotesize
        	{Note} -- Upper limits are reported as the sum of the 97.5th percentiles (3$\sigma$ for a Gaussian distribution) of the posteriors for each velocity component for a given molecule.  Uncertainties on isomer 2 were derived by adding the uncertainties of the individual components in quadrature.\\
        	$^\dagger$No laboratory frequencies available. \\
        \end{minipage}
    \label{tab:ncols}
\end{table}

\begin{figure}[htb!]
    \centering
    \includegraphics[width=\columnwidth]{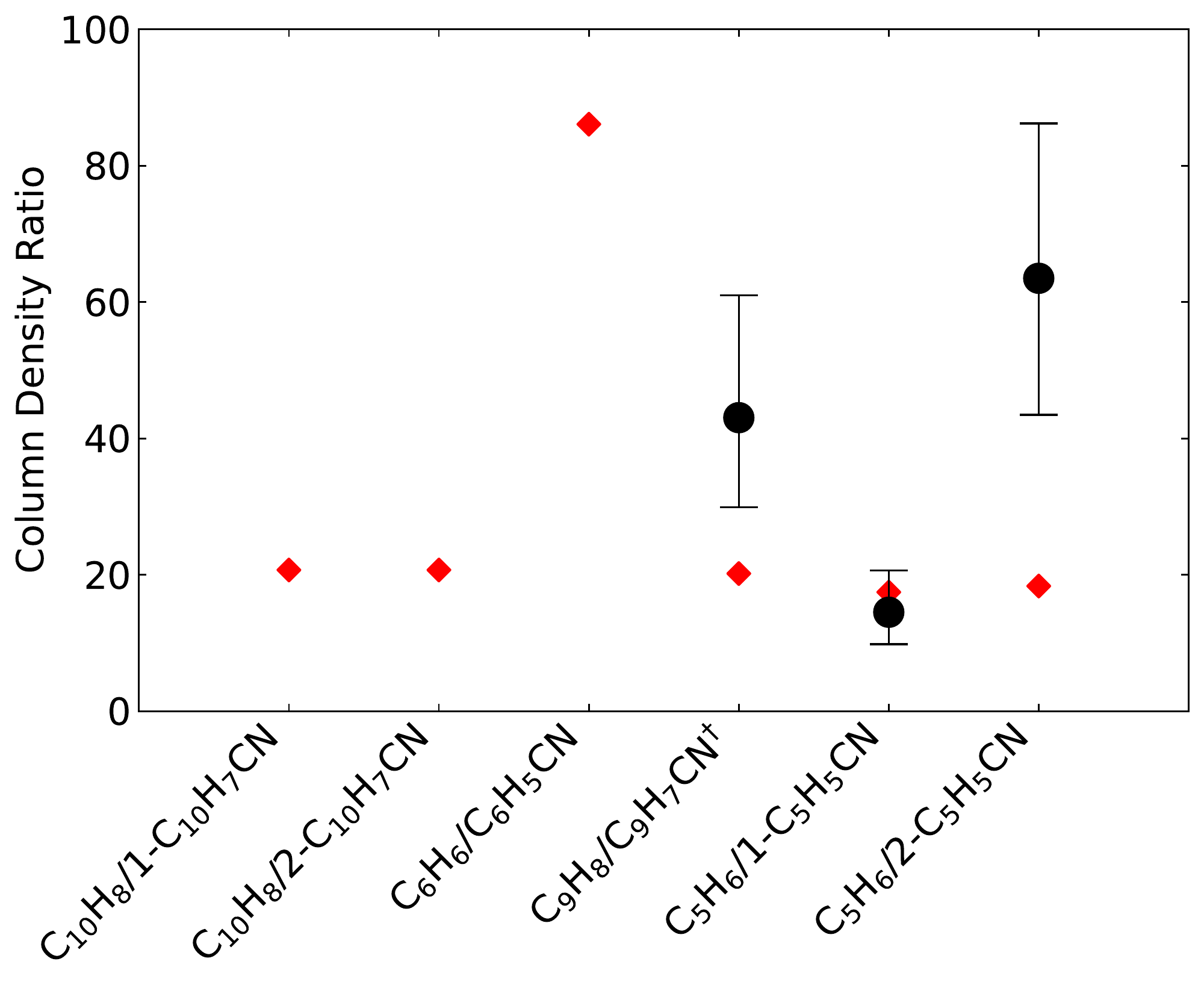}
    \caption{Ratios of the column density of pure hydrocarbon species to their \ce{-CN} substituted counterparts as predicted by our chemical models (red diamonds; \citealt{Burkhardt:2021:L18}) compared to those measured by the GOTHAM and QUIJOTE surveys (black circles with error bars).  Table~\ref{tab:columns} provides the values used and specific references. $^{\dagger}$For \ce{C9H7CN}, the measured value takes only 2-\ce{C9H7CN} into account while the model predicted value includes all the isomers.}
    \label{fig:comparison}
\end{figure}

Finally, Fig.~\ref{fig:comparison} shows the ratios of the column densities of several pure hydrocarbon species to their \ce{-CN} substituted counterparts as measured by observations in TMC-1 and as predicted by the chemical models of  \citet{Burkhardt:2021:L18} {and \citet{Siebert:2022:21}}.  The measured values were obtained by taking the ratio of the reported best-fit total column densities, while the uncertainties were obtained by taking the ratio of the extremes of the errors on those individual measurements.  For example, for the \ce{C9H8}/2-\ce{C9H7CN} ratio, the reported column densities with errors are 9.04$^{+0.96}_{-0.96}$ and 0.210$^{+0.060}_{-0.046}\times10^{12}$\,cm$^{-2}$.  The location of the black circle corresponds to $9.04/0.210 = 43$.  The upper end of the uncertainty was obtained by taking the upper end of the \ce{C9H8} $N_T$ ($9.04 + 0.96 = 10.0$) and dividing by the lower end of the 2-\ce{C9H7CN} $N_T$ ($0.210 - 0.046 = 0.164$) for a final value of $10.0/0.164 = 61$. 

\section{Discussion}

With the detections of benzonitrile (\ce{C6H5CN}) and the two cyanonaphthalene species (\ce{C10H7CN}), the cyanoindenes (\ce{C9H7CN}) were appealing targets to search for.  The detection of 2-\ce{C9H7CN} not only expands the number of individual PAH species detected in space to four, but it offers the first opportunity for a direct observational comparison of a \ce{-CN} substituted PAH to its pure hydrocarbon counterpart. 

Of the possible isomers of \ce{C9H7CN}, we detect only 2-\ce{C9H7CN}.  While 2-\ce{C9H7CN} is the lowest energy species (the next highest being Isomers 4 and 7, roughly 3.4 kJ/mol {(409 K)} higher in energy; see Table~\ref{tab:calculations}), it is not immediately obvious that this should be the only controlling factor.  The reaction pathway currently in astrochemical models for the formation of cyanoindene from \ce{C9H8} is by direct attack by \ce{CN} radical resulting in H-atom loss (Eq.~\ref{rx}; c.f. \citealt{Doddipatla:2021:eabd4044}):
\begin{equation}
    \ce{C9H8 + CN -> C9H7CN + H}.
    \label{rx}
\end{equation}
We note that the model does not distinguish between the possible isomeric forms of cyanoindene, however.

If the reaction between CN and indene in \citet{Doddipatla:2021:eabd4044} is analogous to the production of benzonitrile from benzene via Eq.~\ref{rx_bz}:
\begin{equation}
    \ce{C6H6 + CN -> C6H5CN + H}
    \label{rx_bz}
\end{equation}
then we would expect that attack at any of the double bonds to be efficient and barrierless \citep{Balucani:1999:7457,Cooke:2020:L41}.  It has been shown both experimentally and theoretically that addition–elimination reactions between CN and unsaturated hydrocarbons do not have barriers in their entrance channel and proceed via addition complexes \citep{Carty:2001:310,Sims:1993:461,Woon:1997:204}.  Indeed, that all isomers save 1 and 3 are detected with comparable abundance in the laboratory suggests low site-specificity, albeit in a high-energy, high-density discharge source.  With the exception of the $sp^3$-hybridized carbon at position 1, it is not unreasonable in such an environment to assume a non-zero branching fraction into the different isomers could occur based on the angle of approach of the \ce{CN} radical relative to $sp^2$-hybridized carbons in \ce{C9H8}.

For asymmetrical unsaturated molecules, like indene, however, the site of addition can often dictate the product-branching ratios. In general, the radical will attack at the least substituted position of the unsaturated bond, forming the more stable radical intermediate; however, this is not always the case, as has been observed for propene, where the position of initial CN attack does not determine the product distribution \citep{Huang:2009:12675}. For some addition reactions, optimal orientation of the reactants for the long-range interaction results in a van der Waals potential minimum, which will influence the reaction kinetics and product distribution \citep{CheikhSidEly:2013:12155,Georgievskii:2007:3802}. 

{Our current analysis cannot constrain the other isomers to be significantly less abundant than 2-cyanoindene.  Were further observations to indeed show that they are less abundant, that might then suggest that there could be a degree of thermodynamic control.} Whether or not all of the thermodynamically allowed products can be accessed will depend on the lifetime of the addition complex and the energy barriers to forming products. If the complex is long-lived, energy transfer within the adduct will likely result in a statistical product distribution. It is unlikely that the complex will be collisionally stabilized under TMC-1 conditions. 
If the lifetime is very short, the complex cannot explore all regions of phase space and the products will be controlled by dynamics, i.e. will not be statistical in nature.

{To investigate this further, we have calculated the potential energy surface for the CN addition channels in the CN + \ce{C9H8} reaction (see Appendix C).} Comparing to similar reactions of CN with aromatic hydrocarbons, for example CN + benzene \citep{Woon:2006:67}, CN + toluene \citep{Messinger:2020:7950}, and CN + styrene \citep{Landera:2013:7251}, we expected that the reaction would occur via CN addition to a double bond, either in the 6- or 5-membered ring. The addition proceeds via radical intermediates that can decompose to products or perhaps rearrange to other isomers that can decompose to different products. Determining the product distributions will require a multiwell treatment of the interconnected minima and products. 

{One may also speculate whether CN addition to the double bond in 5-membered ring is favored in these types of systems.} Considering that the additions to each of these sites are barrierless and rapid, the simplest assumption one can make is that they are equally likely. {Our calculations show that} the intermediates formed from addition to the double bonds {(with the exception of the 3-position)} are resonance stabilized radicals {(shown in Figure C3 and C4). The relative energies of the CN-addition complexes are given in Table C2.} Addition to double bonds in the aromatic 6-membered ring is less favorable thermodynamically as the radical intermediates result in loss of the aromaticity. Addition at the 2-position {in the 5-membered ring forms the most stable intermediate of the six possible addition complexes. Addition at this site} forms a benzylic radical intermediate, which may be the reason that is is especially stable. In contrast, addition to the 3-position results in a radical intermediate that cannot be resonance stabilized as the radical site is now displaced from the $\pi$ system of the aromatic ring. More work is required to reduce the uncertainty on the abundances of the other cyanoindene isomers. Since most of their upper limits fall above the observed column density of 2-cyanoindene, we cannot be certain that this isomer is formed preferentially in TMC-1 and more observational evidence is needed. If 2-cyanoindene is indeed the only isomer detected, this would suggest that reaction at the 2-position is kinetically favored or the initial addition complexes have efficient isomerization pathways that can re-orient to form the 2-\ce{C9H7CN} product. Grain-surface pathways might also present a mechanism for re-arrangement into a thermodynamically favored interaction prior to product formation.  Further quantum chemical work is needed to explore the full reaction landscape.

Regardless, we are now able to directly compare the ratio of a pure hydrocarbon PAH to its \ce{-CN} substituted counterpart (hereafter `HC:CN') in observations versus those predicted by our astrochemical models \citep{Burkhardt:2021:L18}.  As we have described before, our models \citep{McGuire:2021:1265,Burkhardt:2021:L18} as well as others \citep{Cernicharo:2021:L9}, usually fail to reproduce the absolute observed abundances of PAHs and other large cyclic species, both purely hydrocarbon and \ce{-CN} derivatives, by several orders of magnitude or more.  As nearly all of the \ce{-CN} species are formed in these models by direct attack by CN radical as shown in, or analogous to, Eqs.~\ref{rx}~\&~\ref{rx_bz}, we suspect the failure of the models is due to incomplete reaction networks leading to the under-production of the pure hydrocarbon species.  While these networks can be improved using the results of ongoing laboratory experiments (see, e.g. \citealt{Doddipatla:2021:eabd4044}) or quantum chemical calculations, many of the pure hydrocarbon species lack a permanent dipole.  As a result, we have no observational constraints on their abundances with which to compare and constrain the updated networks.

One way around this problem is through the observation of the \ce{-CN} derivatives, but this only serves as an effective proxy if we have at least a decent understanding of the expected HC:CN ratio.  As shown in Fig.~\ref{fig:comparison}, the HC:CN ratio predicted by astrochemical models for a range of species pairs for which at least one of the molecules is detected in TMC-1 is relatively consistent at a value of $\sim$20.  We note again that the model results presented here from \citet{Burkhardt:2021:L18} did not distinguish between cyanoindene isomers.  Thus, a full comparison of the ratios of \ce{C9H8}/\ce{C9H7CN} shown in Fig.~\ref{fig:comparison} for the observation and model results would require either a re-working of the model to account for the branching into different isomers (thus increasing the modeled \ce{C9H8}/\ce{C9H7CN} ratio), or the adjustment of the observations to include upper limits of the other isomers (thus lowering the observed \ce{C9H8}/\ce{C9H7CN} ratio).  Either scenario is beyond the scope of the present work, but would actually bring the two values closer into agreement than present.

Of the ratios shown in Fig.~\ref{fig:comparison}, the outlier, the ratio of \ce{C6H6}/\ce{C6H5CN}, still differs only by a factor of $\sim$4 from the others.  This discrepancy is in fact due to the fact that our astrochemical models use the experimentally determined value for {the rate coefficient of}  Reaction~\ref{rx_bz} of 5.4$\times10^{-10}$\,cm$^{3}$\,s$^{-1}$ determined at 15\,K by \citet{Cooke:2020:L41}, whereas the other hydrocarbon + CN reactions use the values of (1.0$-$1.5)$\times10^{-10}$\,cm$^{3}$\,s$^{-1}$ \citep{Burkhardt:2021:L18}. Our results may suggest that this rate is indeed also too low, and we suggest that these reactions would be excellent targets for follow-up experimental studies.  Until then, however, we have no compelling justification to alter the rates in the models.  Nevertheless, the difference in derived ratios remains, by the standards of astrochemical models and observations, quite small.

This consistency is, again, due largely to the similar formation reaction pathways and rates used to produce these species.  Prior observations of the HC:CN ratio for cyclopentadiene and its two CN-derivatives agree with the model predictions to within a factor of $\sim$2--5.  With the detection of 2-\ce{C9H7CN}, we can now posit that that agreement extends to the PAHs as well, and that \ce{-CN} derivative species can indeed be used as an excellent observational proxy for their hydrocarbon counterparts for constraining models, at least within a factor of a few.  

\section{Conclusions}

We report the first interstellar detection of 2-cyanoindene, 2-\ce{C9H7CN}, with a significance of 6.3$\sigma$ in GOTHAM observations of TMC-1.  Of seven potential structural isomers, only the lowest energy form was detected with any significance.  Though chemical models often underpredict the observed abundances of PAH molecules or their \ce{-CN} derivatives in TMC-1, this detection shows that, even so, the calculated ratios of the abundances of the pure hydrocarbon to CN-substituted forms are likely still accurate to within a factor of $\sim$2--5. Thus, combined with existing laboratory and computational studies, we conclude that cyano-PAHs are excellent observational proxies for their pure hydrocarbon counterparts.  Further computational or experimental work is required to fully explain the detection of only a single isomer of \ce{C9H7CN}.

\begin{acknowledgments}
{We gratefully acknowledge support from NSF grants AST-1908576 and AST-2205126.} I.R.C. acknowledges support from the University of British Columbia and the Natural Sciences and Engineering Research Council of Canada (NSERC). The National Radio Astronomy Observatory is a facility of the National Science Foundation operated under cooperative agreement by Associated Universities, Inc. The Green Bank Observatory is a facility of the National Science Foundation operated under cooperative agreement by Associated Universities, Inc.

\textbf{Statement of Efforts.}  All authors contributed to the writing and editing of the manuscript.  M.L.S performed the data reduction and calibration of the DR4 dataset.  P.B.C. performed laboratory measurements. A.M.B. provided astrochemical modeling results.  I.R.C., R.A.L., and C.X. aided in the analysis of the observational data. C.X. designed and performed the observations of Project code: GBT21A-414 and developed the scripts for analysis and corner-plot generation. K.L.K.L. performed the quantum chemical calculations and performed laboratory measurements. E.M. performed and reduced the VLA flux density calibration measurements. {D.G. performed quantum chemical calculations for CN addition to rings.}  C.N.S., I.R.C., E.H., {D.G.,} and P.B.C. contributed to the discussions of formation branching ratios.  B.A.M. performed the analysis of the observational data and designed the project.
\end{acknowledgments}

\bibliography{bam_references}
\bibliographystyle{aasjournal}

\appendix

\renewcommand\thefigure{\thesection\arabic{figure}}   
\renewcommand\thetable{\thesection\arabic{table}}    

\setcounter{figure}{0}    
\setcounter{table}{0} 

\section{MCMC Analysis Results}

\begin{figure}
    \centering
    \includegraphics[width=\textwidth]{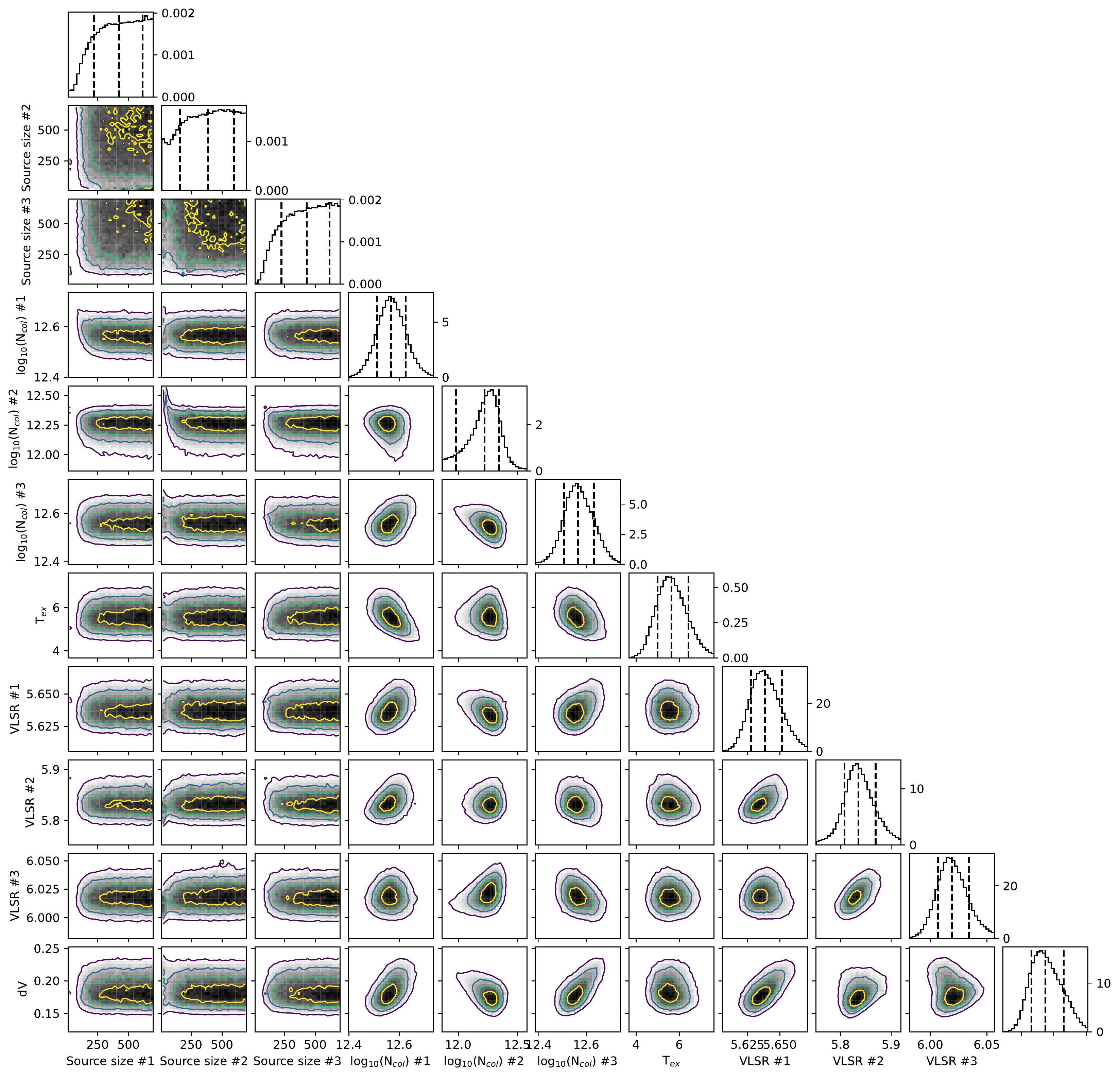}
    \caption{Corner plot for \ce{C9H8} showing parameter covariances and marginalized posterior distributions for the \ce{C9H8} MCMC fit. 16$^{th}$, 50$^{th}$, and 84$^{th}$ confidence intervals (corresponding to $\pm$1 sigma for a Gaussian posterior distribution) are shown as vertical lines. }
    \label{fig:indene_corner}
\end{figure}

\begin{figure}
    \centering
    \includegraphics[width=\textwidth]{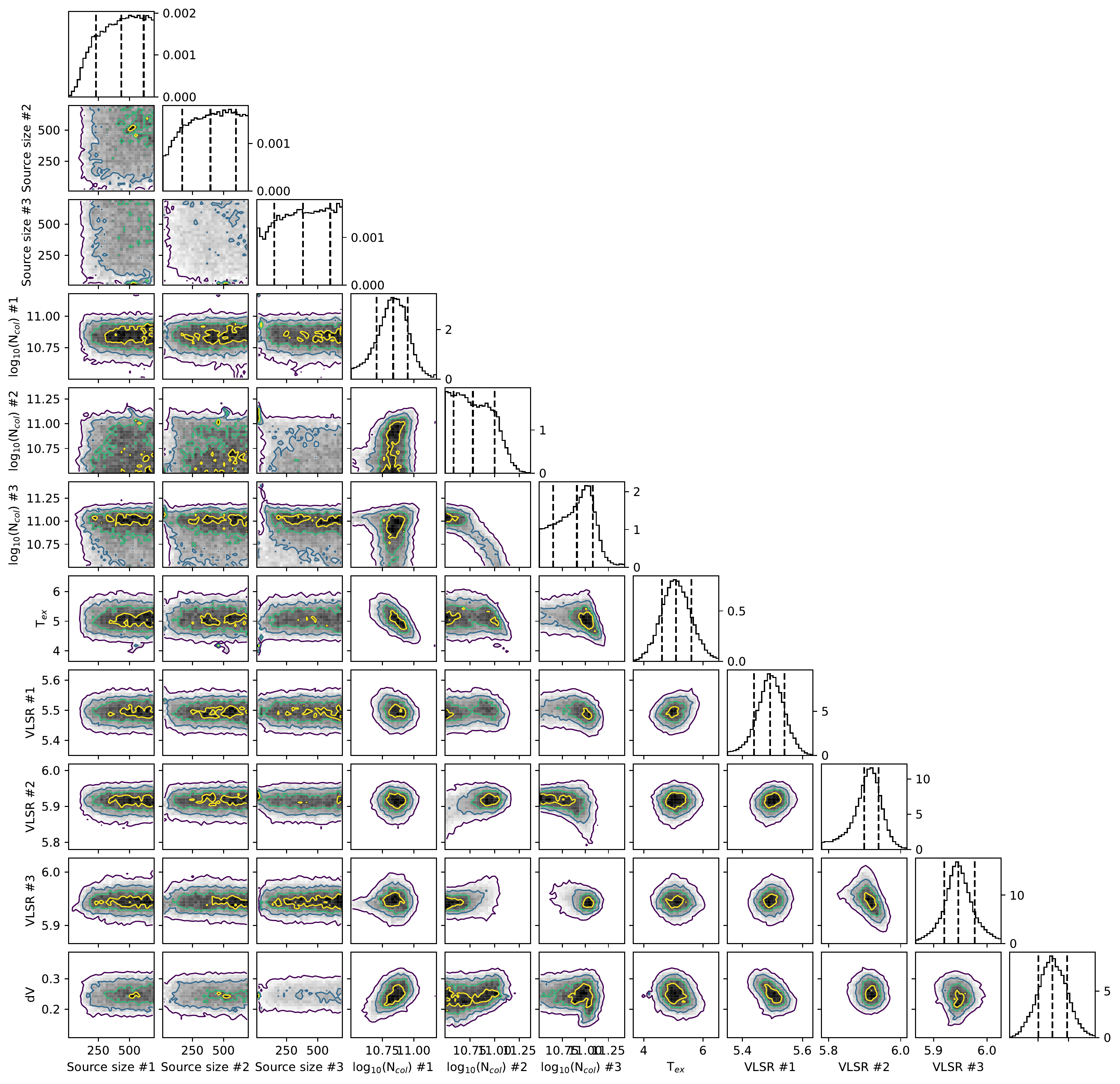}
    \caption{Corner plot for 2-\ce{C9H7CN} showing parameter covariances and marginalized posterior distributions for the 2-\ce{C9H7CN} MCMC fit. 16$^{th}$, 50$^{th}$, and 84$^{th}$ confidence intervals (corresponding to $\pm$1 sigma for a Gaussian posterior distribution) are shown as vertical lines. }
    \label{fig:indene_cn_corner}
\end{figure}

The corner plots resulting from the analysis of \ce{C9H8} and 2-\ce{C9H7CN} are shown in Fig.~\ref{fig:indene_corner} and Fig.~\ref{fig:indene_cn_corner}, respectively.

\clearpage
\section{Observed Column Densities of Hydrocarbon Species and Their \ce{-CN} Substituted Analogs}

Table~\ref{tab:columns} provides the column densities of those species plotted in Fig.~\ref{fig:comparison} observed in TMC-1.

\begin{table}[h!]
    \centering
    \caption{Column densities of species plotted in Fig.~\ref{fig:comparison} observed in TMC-1.}    
    \begin{tabular}{c c c}
    \toprule
    Molecule        &   Column Density          &   Reference                   \\
                    &    (10$^{12}$ cm$^{-2}$)  &                               \\
    \ce{C5H6}       &   12$^{+3}_{-3}$           &   \citet{Cernicharo:2021:L15} \\
    1-\ce{C5H5CN}   &   0.827$^{+0.09}_{-0.10}$   &   \citet{Lee:2021:L2}         \\
    2-\ce{C5H5CN}   &   0.189$^{+0.018}_{-0.015}$ &   \citet{Lee:2021:L2}         \\
    \ce{C9H8}       &   9.04$^{+0.96}_{-0.96}$    &   This Work                   \\
    2-\ce{C9H7CN}     &   0.210$^{+0.060}_{-0.046}$ &   This Work                   \\
    \midrule
    \bottomrule
    \end{tabular}
    \label{tab:columns}
\end{table}

\section{Quantum Chemical Calculations for CN addition to the indene ring}

Table~\ref{tab:complex-energies} lists the relative energies of the complexes formed after CN addition to the indene ring. It is noted that the complex formed by CN addition at the carbon position 2 is the most stable among the 6 possible addition complexes.

\begin{table}[h!]
    \centering
    \caption{Relative energies, provided in kJ/mol, of the CN-addition complexes calculated at (U)B3LYP-D3/def2-TZVP.}    
    \begin{tabular}{c c c}
    \toprule
    CN Addition Complex        &   Relative Energy (kJ/mol)      \\
    \midrule
    2-\ce{C9H8CN}    &    -263.7  \\
    3-\ce{C9H8CN}      &   -217.9    \\
    4-\ce{C9H8CN}    &   -187.5 \\
    5-\ce{C9H8CN}    &   -174.0   \\
    6-\ce{C9H8CN}    &   -191.9  \\
    7-\ce{C9H8CN}    &   -174.9 \\
    \bottomrule
    \end{tabular}
    \label{tab:complex-energies}
\end{table}

\begin{figure}
    \centering
    \includegraphics[width=\textwidth]{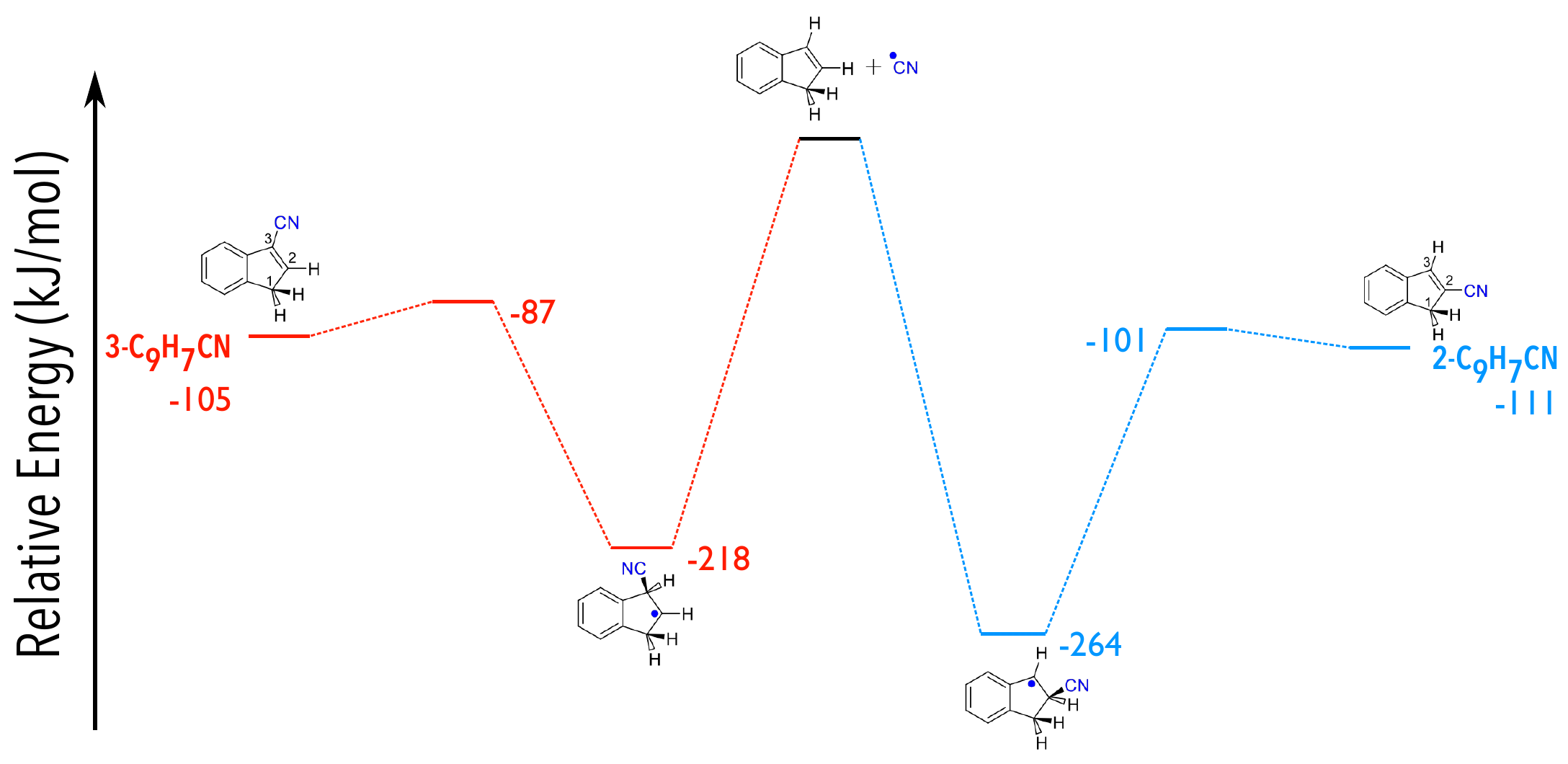}
    \caption{The potential energy surface ((U)B3LYP-G3/def2-TZVP) for the CN addition--H elimination channels for the reaction between CN and indene. Two possibile channels occur via CN addition at carbon positions 2 and 3 for the 5-membered ring.}
    \label{fig:pes-CN-5ring}
\end{figure}

\begin{figure}
    \centering
    \includegraphics[width=\textwidth]{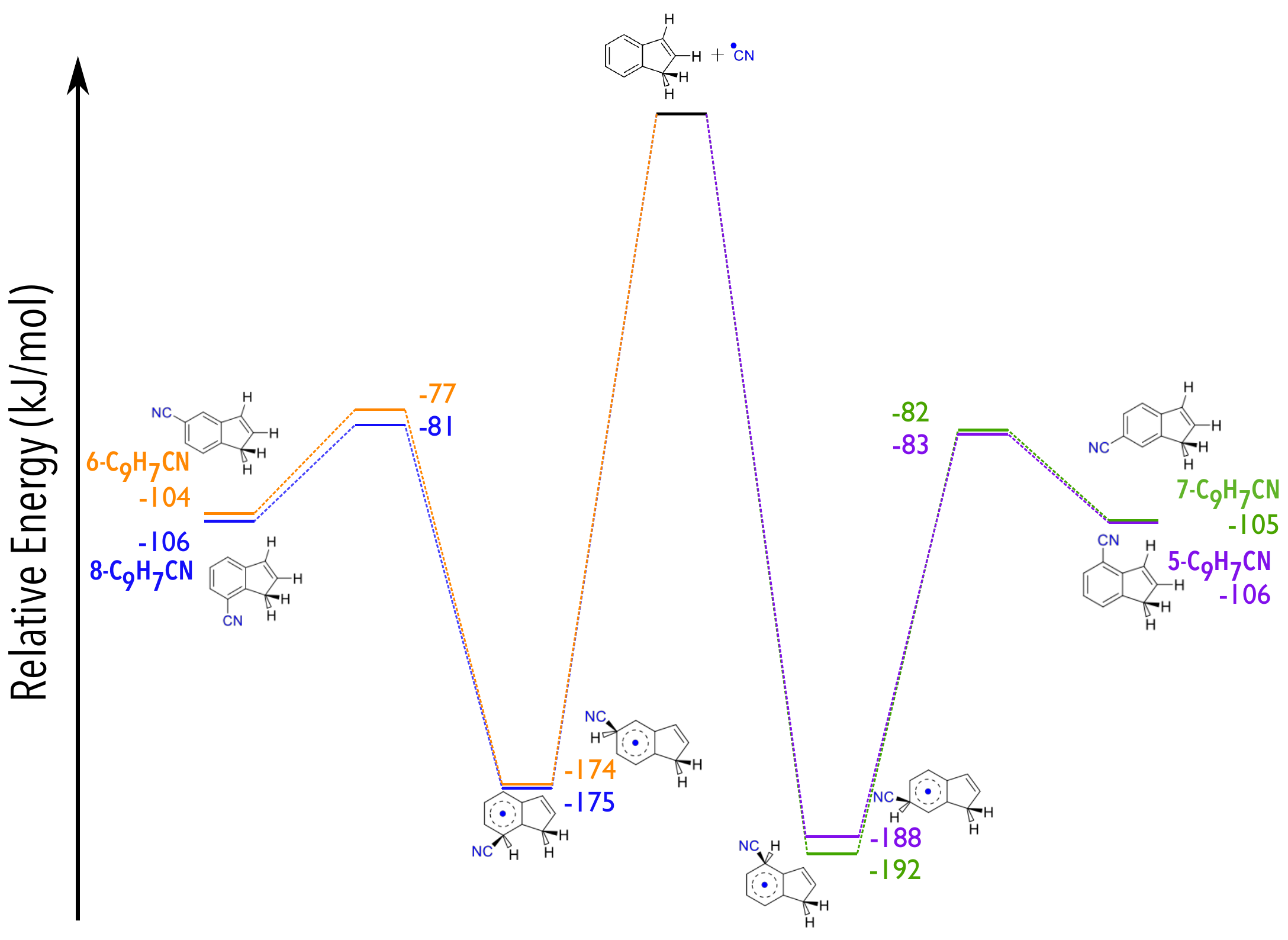}
    \caption{The potential energy surface ((U)B3LYP-G3/def2-TZVP) for the CN addition--H elimination channels for the reaction between CN and indene. Four possibile channels occur via CN addition at carbon positions 5, 6, 7 and 8 for the 6-membered ring.}
    \label{fig:pes-CN-6ring}
\end{figure}

Figures \ref{fig:pes-CN-5ring} and \ref{fig:pes-CN-6ring} show the complete potential energy surface obtained for the CN addition channels for \ce{CN} + indene.  The quantum calculations were performed using Gaussian 16 software \citep{Gaussian16}. All the species, including the addition complexes and transition states, were optimized at the (U)B3LYP-G3/def2-TZVP level, and zero-point corrected energies were calculated for each. In addition, intrinsic reaction coordinate (IRC) calculations were performed at (U)B3LYP-G3/def2-TZVP to determine the minimum energy path that the transition states followed to confirm the connection between the appropriate reactants and products. 

\end{document}